\documentclass[preprint,aps,12pt,preprintnumbers,eqsecnum,nofootinbib,superscriptaddress]{revtex4}
\usepackage{graphicx}
\usepackage{amssymb,amsmath}
\def\lag{{\mathcal{L}}}

\def\be{\begin{equation}}
\def\ee{\end{equation}}
\def\beq{\begin{eqnarray}}
\def\eeq{\end{eqnarray}}
\def\f{\frac}
\def\lf{\left}
\def\rh{\right}

\begin{document}

\title{Constraints on the Lee-Wick Higgs Sector}

\author{Christopher D. Carone}
\email[]{cdcaro@wm.edu}

\author{Reinard Primulando}
\email[]{rprimulando@wm.edu}
\affiliation{Department of Physics, College of William \& Mary, Williamsburg, VA 23187-8795}

\date{August, 2009}

\begin{abstract}
Lee-Wick partners to the Standard Model Higgs doublet may appear at a mass scale that
is significantly lower than that of the remaining Lee-Wick partner states.  The relevant effective 
theory is a two-Higgs doublet model in which one doublet has wrong-sign kinetic and mass terms.  
We determine bounds on this effective theory, including those from neutral B-meson mixing, 
$b \rightarrow X_s \gamma$, and $Z \rightarrow b \overline{b}$.  The results differ from 
those of conventional two-Higgs doublet models and lead to meaningful constraints on the 
Lee-Wick Higgs sector.
\end{abstract}

\maketitle

\section{Introduction}\label{sec:intro}

The Lee-Wick Standard Model (LWSM) presents a novel solution for
addressing the hierarchy problem in the Standard Model~\cite{GOW1}.  For every
Standard Model field, a gauge-invariant, higher-derivative (HD) kinetic 
term is introduced so that propagators fall off more quickly with momentum.
Although such terms include higher-derivative interactions as well, 
a power-counting exercise shows that the resulting theory is no more 
than logarithmically divergent~\cite{GOW1}.  The dependence of the Higgs boson
mass squared on any ultraviolet physics is no worse in the LWSM than 
it is, for example, in the Minimal Supersymmetric Standard Model (MSSM).

The presence of HD quadratic terms leads to the presence of additional
poles in the two-point function of each field.  Using an auxilliary
field method (that we review in the next section), it is possible to 
recast the original LWSM Lagrangian in terms of one without HD terms, but with 
additional fields that correspond to the LW partner states~\cite{GOW1}.  In this 
formulation, all interactions in the Lagrangian have mass dimension no greater than four.  
We refer to this as the LW form of the theory. The enhanced convergence of 
the LWSM in the HD form is reproduced in the LW form via cancellations
between Feynman diagrams involving Standard Model particles and those 
involving their LW partners.  Such diagramatic cancellations are reminiscent 
of the situation in the MSSM, but differ in a fundamental way: LW particles
have the same spin as their Standard Model partners.  The cancellation of 
quadratic divergences is a consequence of the signs of the LW kinetic
and mass terms, which are opposite to those of ordinary particles.  

It is natural to question the consistency of a theory that includes physical
states with ``wrong-sign" kinetic and mass terms.  These states have negative
norms, so that the free Hamiltonian is bounded from below.  The presence of 
eigenstates of the Hamiltonian with real eigenvalues and negative norms, however, 
can lead to a violation of unitarity.  Lee and Wick~\cite{LW1,LW2} showed 
long ago that unitary can be preserved in such a theory if the negative norm states have 
non-vanishing decay widths, and hence are eigenstates of the Hamiltonian with 
complex eigenvalues.  The $S$ matrix constructed out of the eigenstates of the 
Hamiltonian with real eigenvalues excludes these states and is unitary.  Lee and 
Wick~\cite{LW1,LW2}, as well as Cutkosky {\em et al.}~\cite{CLOP} showed how the pole 
prescription in perturbation theory is modified so that the correct $S$ matrix is 
produced, and no violation of unitarity was found in any explicit higher-order calculation 
that the authors considered.  While this construction of the $S$ matrix leads to a 
violation of causality at a microscopic level, no logical paradoxes have been shown to 
arise at macroscopic scales~\cite{LW1,LW2,coleman,causal}.  More recently, unitarity of 
longitudinal gauge boson scattering amplitudes in the LWSM has
been demonstrated~\cite{wwscat}, a result that is not obvious given that LW vector bosons masses do 
not arise via spontaneous symmetry breaking.  In summary, every explicit calculation
in LW theories, including nonperturbative studies~\cite{nonperturb}, has supported the 
consistency of these theories.  This motivates phenomenological 
studies~\cite{collide,dandw,ewcons,CL1,hdlw,lwloop,lwu,lwht,lwother} of
the LWSM as a viable solution to the hierarchy problem.

Recent work on the LWSM has included studies of collider signals~\cite{collide}, 
flavor-changing~\cite{dandw} and electroweak precision constraints~\cite{ewcons,CL1}, 
higher-derivative generalizations~\cite{hdlw}, running of couplings and unification~\cite{lwloop,lwu}, 
and LW theories at high-temperature~\cite{lwht}.  If the LW particles are assumed 
to have a common mass, $M_{LW}$, then electroweak constraints typically require this 
scale to be above $\sim 5$~TeV.  However, as pointed out in Ref.~\cite{CL1}, the spectrum of 
LW particles need not be degenerate.  The LW partners to the Higgs boson, the top quark, 
the left-handed bottom quark and the SU(2) gauge bosons give the greatest contributions 
to the cancellation of the Higgs boson quadratic divergence and must be present at or 
below the TeV scale.  The remaining LW partners could appear around $10$~TeV without 
requiring substantial fine tuning in the Higgs bare mass.  The electroweak
constraints on the effective theory with this minimal LW particle content was studied 
in Ref.~\cite{CL1}, where it was noted that the LW-mass scale for the Higgs sector $m_{\tilde{h}}$ 
was only weakly constrained.  While the LW gauge and fermion partners were again forced into the
multi-TeV range, $m_{\tilde{h}}$ could be $O(100)$~GeV without running afoul of the bounds.
This result suggests another possible hierarchy in the LW particle spectrum: the LW partners 
to the Standard Model Higgs doublet could be well below $1$~TeV, while the remaining LW 
states could be substantially heavier than the LW Higgs.

What is interesting about this effective theory is that it is similar to the often-studied
two-Higgs doublet (2HD) extensions of the Standard Model.  However, the sign difference in the 
LW kinetic and mass terms leads to sign changes in the LW Higgs propagator as well as in
the interaction vertices that originate in the kinetic terms by gauge invariance.  Sign 
differences in specific Feynman diagrams change the theoretical predictions for a number
of one-loop processes, so that the resulting bounds on the scale $m_{\tilde{h}}$ cannot be
simply inferred from the 2HD results; a numerical reanalysis is required.  The purpose of this
paper is to begin this task, by considering the bounds from neutral B-meson mixing (for both the
$B_d$ and $B_s$ mesons), the decay $b \rightarrow X_s \gamma$ and the decay 
$Z \rightarrow b \overline{b}$.  The bounds on the LW Higgs sector substantially supercede 
those that appear in Ref.~\cite{CL1} and are relevant in determining the parameter space that 
might be explored in future collider experiments.

Our paper is organized as follows:  In Section~\ref{sec:two}, we review the construction of the 
LW Higgs sector and establish our conventions.  In Section~\ref{sec:three} we determine the bounds 
on the charged LW Higgs from B-meson mixing , $b \rightarrow X_s \gamma$ and 
$Z \rightarrow b \overline{b}$.  In Section~\ref{sec:four} we consider the constraints that 
are implied by these results on the neutral LW Higgs states, and in the final section we 
summarize our conclusions.

\section{Higgs Sector of the LWSM} \label{sec:two}
The Higgs sector of the LWSM is given by the Lagrangian
\be \label{eq:laghd}
\mathcal L_{HD} = (D_\mu \hat H)^{\dagger} (D^\mu \hat H) 
-\f{1}{m^2_{\tilde h}} (D_\mu D^\mu \hat H)^{\dagger} (D_\nu D^\nu \hat H) - V(\hat H),
\ee
where the hat indicates that the field is defined in the HD theory.  Since 
the LW gauge bosons are decoupled from the effective theory of interest, the covariant 
derivative is given by
\be
D_\mu = \partial_\mu + ig_2 W_\mu^a T^a + ig_1 B_\mu Y.
\ee
where $W^a_\mu$ and $B_\mu$ are the ordinary SU(2)$_W$ and U(1)$_Y$ gauge fields, respectively.  Note 
that the generators are normalized such that $\mbox{Tr }T^a T^b = 1/2$ and $Y \hat{H} = 1/2 \,\hat{H}$. 
The potential $V(\hat H)$ takes the form
\be
V(\hat H) = \f{\lambda}{4} \lf(\hat H^\dagger \hat H -\f{v^2}{2}\rh)^2.
\ee
In order to eliminate the higher-derivative term in Eq.~(\ref{eq:laghd}), we construct an equivalent
Lagrangian using an auxiliary field $\tilde H$~\cite{GOW1}:
\be\label{eq:laf}
\mathcal L_{AF} = 
(D_\mu \hat H)^{\dagger} (D^\mu \hat H) + (D_\mu \hat H)^\dagger (D^\mu \tilde H) 
+ (D_\mu \tilde H)^\dagger (D^\mu \hat H) + m_{\tilde h}^2\tilde H^\dagger \tilde H  - V(\hat H).
\ee
One recovers the Lagrangian in Eq.~(\ref{eq:laghd}) by substituting the equation of motion for the 
auxiliary field into Eq.~(\ref{eq:laf}) and integrating by parts.  The kinetic terms are diagonalized
by the field redefinition $\hat H = H - \tilde H$, yielding
\be \label{eq:GaugeHiggs} 
\mathcal L = (D_\mu H)^{\dagger} (D^\mu H) 
- (D_\mu \tilde H)^{\dagger} (D^\mu \tilde H) + m_{\tilde h}^2\tilde H^\dagger\tilde H- V (H - \tilde H).
\ee
The higher-derivative term has been eliminated at the expense of introducing the LW field $\tilde H$
which has wrong-sign kinetic and mass terms.  

The last two terms in Eq.~(\ref{eq:GaugeHiggs}) are extremized when the field $H$ acquires a 
vacuum expectation value.  In unitary gauge, one can write~\cite{GOW1}
\be
H = \begin{pmatrix}
0 \\ \f{v+h}{\sqrt{2}}
\end{pmatrix}, \;\;\;
\tilde H = \begin{pmatrix}
\tilde h^{+} \\ \f{\tilde h + i \tilde P}{\sqrt{2}}
\end{pmatrix}.
\ee
where $v \approx 246$~GeV sets the electroweak scale.  We will refer to $h$ the ordinary Higgs field,
and $\tilde h$, $\tilde P$, and $\tilde h^+$ as the LW scalar, pseudoscalar and charged Higgs fields,
respectively.  The Higgs field masses are determined by
\be
\mathcal L_{\textrm{mass}} = -\f{\lambda}{4}v^2(h-\tilde h)^2 
+ \f{m_{\tilde h}^2}{2} ( \tilde h \tilde h + \tilde P \tilde P + 2 \, \tilde h^+ \tilde h^-),
\ee
which shows mixing between the ordinary Higgs scalar and its LW partner.  One can diagonalize 
the scalar mass matrix without altering the form the kinetic terms via a 
symplectic transformation:
\begin{equation} \label{symprot}
\left( \begin{array}{c} h \\ \tilde h \end{array} \right) = \left(
\begin{array}{cc} \cosh \theta & \sinh \theta \\ \sinh \theta & \cosh
\theta \end{array} \right) \left( \begin{array}{c} h_0 \\ \tilde h_0
\end{array} \right) \, ,
\end{equation}
where subscript $0$ indicates a mass eigenstate.  The mixing angle $\theta$ satisfies
\begin{equation} \label{tanhform}
\tanh 2\theta = \frac{-2m_h^2/m_{\tilde h}^2}{1-2m_h^2/m_{\tilde h}^2}
= -\frac{2m_{h_0}^2 m_{\tilde h_0}^2}{m_{h_0}^4 +
m_{\tilde h_0}^4} \, ,
\end{equation}
with mass eigenvalues
\be \label{eq:eigenmass}
m_{h_0}^2 = \frac{m_{\tilde h}^2}{2} \left( 1 -
\sqrt{ 1 -\frac{4 m_h^2}{m_{\tilde h}^2} } \right) \,\,\,\,\, \mbox{ and } \,\,\,\,\,
m_{\tilde h_0}^2 = \frac{m_{\tilde h}^2}{2} \left( 1 +
\sqrt{ 1 -\frac{4 m_h^2}{m_{\tilde h}^2} } \right) \,.
\ee
Notice that the eigenstate with ``wrong-sign" kinetic terms is always the heavier of the two.
The LW pseudoscalar $\tilde P$ and LW charged scalar $\tilde h^+$ have the same mass $m_{\tilde h}$.
The masses of the neutral scalars are related to the mass of the pseudoscalar or charged Higgs by
\be \label{eq:massrelation} 
m_{h_0}^2 + m_{\tilde h_0}^2 = m_{\tilde h}^2.
\ee

As in the gauge sector, the LW partners to the SM fermions are decoupled from our effective 
theory. Even assuming realistic LW fermion masses of a few TeV, mixing between ordinary and 
LW fermions is numerically small and can be ignored.  The Yukawa couplings involving $H$ 
and $\tilde{H}$ are then given by~\cite{GOW1}
\begin{eqnarray} \label{eq:Yukawa} 
\delta\lag &=& \frac{\sqrt 2}{v} \sum_i \left[ m_u^i \overline{u^i}_R (H - \tilde H) \epsilon Q_L^i
 - m_d^i \overline{d^i}_R (H^\dagger - \tilde H^\dagger) Q_L^i  \right. \nonumber \\
 &-& \left. m_e^i \overline{e^i}_R (H^\dagger - \tilde H^\dagger) L_L^i
+ \mathrm{ h.c.} \right] ,
\end{eqnarray}
where
\begin{equation}
Q_L = \begin{pmatrix}
u_L \\ V d_L
\end{pmatrix} \, ,\,\,\,\,\,\,\,\,\,\,\, L_L = \left(\begin{array}{c} \nu \\ e_L\end{array}\right) \,,
\end{equation}
$V$ is usual CKM matrix, and the fields are given in the mass eigenstate basis.

\section{Constraining the Charged Higgs} \label{sec:three}

The interaction between quarks and the charged Higgs field in LWSM can be extracted 
from Eq.~(\ref{eq:Yukawa}),
\begin{equation} \label{eq:hff} 
\mathcal{L}_{\tilde{h}^\pm ff} = - \frac{g_2}{\sqrt{2} M_W} \tilde{h}^{+} \sum_{ij} 
\left[ m_u^i \bar{u}^i_R V_{ij} d_L^j + m_d^j \bar{u^i}_L V_{ij} d_R^j \right] + \textrm{ h.c.},
\end{equation}
while the  $\gamma$-Higgs-Higgs and $Z$-Higgs-Higgs couplings follow from Eq.~(\ref{eq:GaugeHiggs}),
\be \label{eq:hhAZ}
\mathcal{L}_{\tilde{h}^+\tilde{h}^-A,Z} = \left(ie A^{\mu}+i\f{g_2\cos2\theta_W}{2\cos\theta_W}Z^\mu\right)
\left(\tilde{h}^- \partial_\mu \tilde{h}^+ - \tilde{h}^+ \partial_\mu \tilde{h}^- \right).
\ee
The analogous couplings in a Two-Higgs-Doublet Model (2HDM) of type II are given by~\cite{Gunion:1984yn}
\begin{equation} \label{eq:HDMhff} 
\mathcal{L}_{h^\pm ff}^{2HDM} = \frac{g_2}{\sqrt{2} M_W} h^{+} \sum_{ij} \left[  
\cot\beta m_u^i \bar{u}^i_R V_{ij} d^j_L
+ \tan\beta m_d^j \bar{u^i}_L V_{ij} d_R^j \right] + \textrm{ h.c.}
\end{equation}
and
\be \label{eq:HDMhhAz}
\mathcal{L}_{h^+h^-A,Z}^{2HDM} = - \left(ie A^{\mu}+i\f{g_2\cos2\theta_W}{2\cos\theta_W}Z^\mu\right)
\left(h^- \partial_\mu h^+ - h^+ \partial_\mu h^- \right).
\ee
By comparing Eqs.~(\ref{eq:hff}) and (\ref{eq:hhAZ}) with Eqs.~(\ref{eq:HDMhff}) and (\ref{eq:HDMhhAz}), 
we see that the charged Higgs interactions in the LWSM mimic those of the type-II 2HDM with 
$\tan\beta =1$, except for overall signs.  Hence, each occurrence of an  $\tilde{h}^+\tilde{h}^- \gamma$,
$\tilde{h}^+\tilde{h}^- Z$ or $\tilde{h}^\pm \overline{q} q$ vertex in a Feynman diagram introduces 
a factor of $-1$ relative to the corresponding result in a type-II 2HDM.  In addition, every charged Higgs
propagator introduces another factor of $-1$, due to the wrong-sign LW quadratic terms.  These
observations allow us to modify the phenomenologically relevant, but sometimes complicated, 
next-to-leading order (NLO) calculations of loop amplitudes in the type-II 2HDM to determine the
numerical bounds on the LW Higgs sector.   The processes that we consider below are ones that
are enhanced by the large top quark Yukawa coupling; one would expect these to provide a reasonable
picture of the allowed parameter space of the effective theory. 

We consider $B_q\bar B_q$ mixing for $q=d$ or $s$, the inclusive decay $B \rightarrow X_s \gamma$ and 
$R_b \equiv \Gamma(Z\rightarrow b \overline{b})/\Gamma(Z\rightarrow \mbox{ hadrons})$ to obtain 
bounds on the charged Higgs mass.   These processes have been evaluated in the type-II 2HDM including 
NLO QCD corrections in Refs.~\cite{urban}, \cite{ciuch,borz,gambino} and \cite{denner,howie}, respectively.  
As described in the previous paragraph, we modify the 2HD model amplitudes to obtain bounds on the mass 
of the LW charged Higgs $\tilde{h}^\pm$.  

\subsection{$B_q \overline{B}_q$ mixing}
\begin{figure}
\centering
\includegraphics[width=75mm,angle=0]{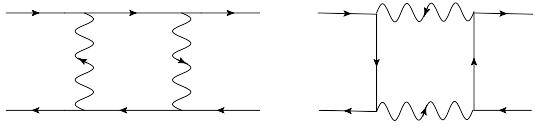}
\caption{SM diagrams for $B\bar B$ mixing. Wavy lines represent $W$ bosons and solid lines represent quarks.}
\label{fig:Box1}
\end{figure}

\begin{figure}
\centering
\includegraphics[width=75mm,angle=0]{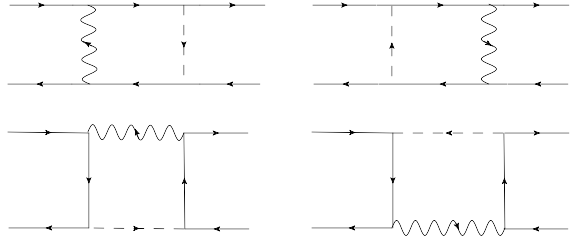}
\caption{New physics diagrams for $B\bar B$ mixing with one $W$ and one charged Higgs exchanged. 
Dashed lines represent charged Higgs fields, wavy lines represent $W$ bosons and solid lines 
represent quarks. These diagrams are proportional with $\cot^2\beta$ in the type-II 2HDM and 
flip overall sign in the LWSM.}
\label{fig:Box2}
\end{figure}

\begin{figure}
\centering
\includegraphics[width=75mm,angle=0]{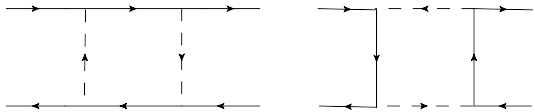}
\caption{New physics diagrams for $B\bar B$ mixing with two charged Higgs particles exchanged. 
Dashed lines represent charged Higgs fields and solid lines represent quarks. These diagrams 
are proportional with $\cot^4\beta$ in 2HDM and and have the same sign in the LWSM.}
\label{fig:Box3}
\end{figure}

Before including the NLO QCD corrections, it is instructive to consider the leading-order (LO) result 
evaluated at the matching scale at which the exotic Higgs physics is integrated out.  This scale is
typically taken to be $m_W$. The mass splitting between $B^0_q$ and $\overline{B^0}_q$ mesons in the 
2HDM of type II is then given by~\cite{hhg}
\be
\Delta m_{B_{2HDM}} = \f{G_F^2}{6\pi^2}m_W^2 \lf|V_{tq}V_{tb}^*\rh|^2\ f_B^2 \hat{B}_{B_q} m_B 
\lf( I_{WW}+\cot^2\beta\ I_{Wh}+\cot^4\beta \  I_{hh} \rh)  \,\,\,.
\ee
Here $I_{WW}$ originates from the pure $W^{\pm}$-exchange Feynman diagrams shown in Fig.~\ref{fig:Box1}, 
$I_{Wh}$ from the single-charged-Higgs-exchange diagrams in Fig.~\ref{fig:Box2}, and $I_{hh}$ from
the pure charged-Higgs-exchange diagrams shown in Fig.~\ref{fig:Box3}. These functions are given by~\cite{hhg}
\beq
I_{WW}&=&\f{x}{4} \left(1+\f{9}{\lf(1-x\rh)}-\f{6}{\lf(1-x\rh)^2} 
- \f{6}{x} \lf(\f{x}{1-x}\rh)^3\ln x\right) \, ,\nonumber \\
I_{Wh}&=&\f{xy}{4} \left[ -\f{8-2x}{(1-x)(1-y)}+\f{6z\ln x}{(1-x)^2(1-z)} 
+\f{\lf(2z-8\rh)\ln y}{(1-y)^2(1-z)} \right] \, ,\nonumber \\
I_{hh}&=&\f{xy}{4} \lf[ \f{\lf(1+y\rh)}{\lf(1-y\rh)^2} + \f{2y\ln y}{\lf(1-y\rh)^3} \rh] \, ,
\label{eq:lofuncs}\eeq
where $x = {m_t^2}/{m_W^2}$, $y={m_t^2}/{m_{\tilde h}^2}$ and $z={x}/{y}={m_{\tilde h}^2}/{m_W^2}$. 

\begin{figure}
\includegraphics[scale=0.75]{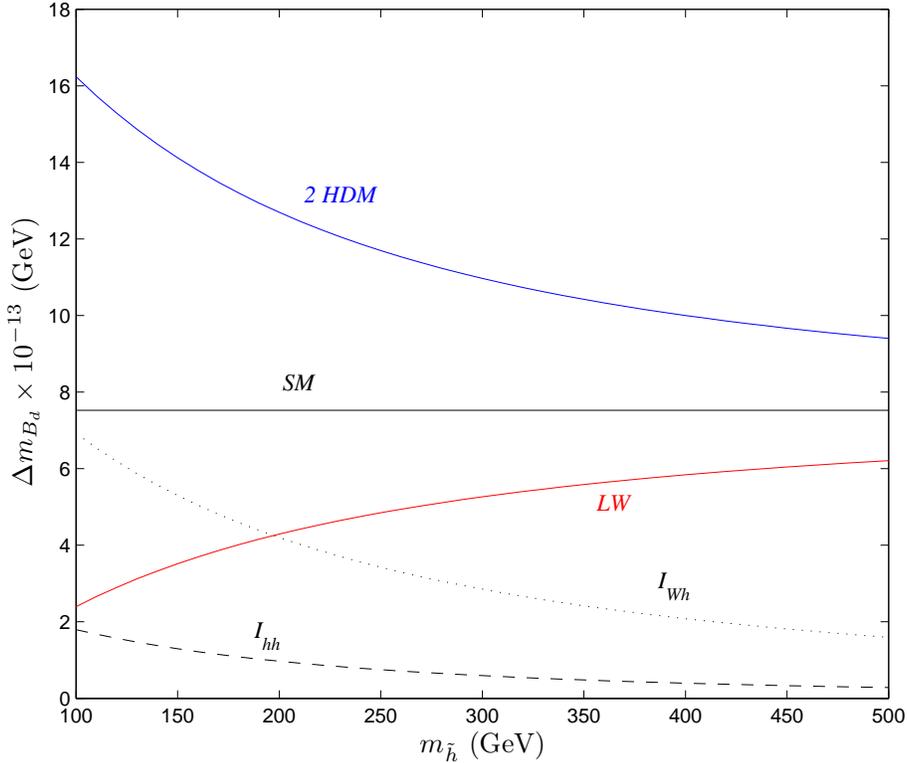}
\caption{The mass splitting $\Delta m_{B_d}$ predicted at LO in the LWSM and in type-II 2HDM 
with $\tan\beta=1$, as functions of the charged Higgs mass. The solid lines represents the
predictions of the 2HDM, the Standard Model and LWSM, as labelled. The dashed line gives 
$\f{G_F^2}{6\pi^2}m_W^2 \lf(V_{tq}V_{tb}^*\rh)^2\ f_B^2 \hat{B}_{B_d} m_{B_d} \times I_{Wh}$ and 
dotted line gives $\f{G_F^2}{6\pi^2}m_W^2 \lf(V_{tq}V_{tb}^*\rh)^2\ f_B^2 
\hat{B}_{B_d} m_{B_d} \times I_{hh}$.}
\label{fig:Igraph}
\end{figure}

To obtain the mass splitting appropriate to the LWSM, we set $\tan\beta = 1$.  We must multiply $I_{Wh}$ 
by $(-1)^3=-1$, which takes into account two $f\bar{f}\tilde{h}^\pm$ vertices and one $\tilde{h}^\pm$ 
propagator; we multiply $I_{hh}$ by $(-1)^6=1$ because there are four $f\bar{f}\tilde{h}^\pm$ vertices 
and two $\tilde{h}^\pm$ propagators.  Therefore, for the LWSM, one finds the LO result
\be
\Delta m_{B_{LWSM}} = \f{G_F^2}{6\pi^2}m_W^2 \lf|V_{tq}V_{tb}^*\rh|^2\ f_B^2 
\hat{B}_{B_q} m_B\lf(I_{WW}-I_{Wh}+I_{hh} \rh) 
\ee
Our numerical values for the particle masses, CKM elements $V_{ij}$, $G_F$, the decay constant $f_B$ and 
the bag factor $\hat{B}_{B_q}$ are given in Appendix~\ref{inputs}.  Using these, we plot the LO
$B_d-\bar{B}_d$ mass splitting in the LWSM and in a type-II 2HDM with $\tan\beta=1$ in 
Fig.~\ref{fig:Igraph}.  The new physics diagrams in the 2HDM give a positive contribution 
to the mass splitting.  In the LWSM, however, the mass splitting receives a negative contribution since
the $I_{Wh}$ term flips sign and dominates over $I_{hh}$.  Since the magnitude of $I_{Wh}$ is comparable
to that of $I_{WW}$, the new physics can significantly alter the Standard Model prediction, leading
to useful bounds on the mass of the charged Higgs when the result is compared to the experimental value.

To do such a comparison, however, we work with the NLO result that includes QCD corrections and 
takes into account running between the matching scale $m_W$ and the scale of the $B$-mesons.
Unlike Eq.~(\ref{eq:lofuncs}), these amplitudes are quite complicated and cannot be summarized in a 
few lines.  However, the approach for modifying them to obtain LWSM results is precisely the 
same as in our simple leading order example.  We use the NLO amplitudes given in Ref.~\cite{urban} for 
our numerical analysis.  Our predictions depend on the bag factor which is the largest source of 
theoretical uncertainty.  We use lattice QCD estimates of the bag factors given in Ref.~\cite{Gamiz:2009ku}: 
$f_B \sqrt{\hat{B}_{B_d}} = 216\pm 15 \textrm{ GeV}$ and $f_B \sqrt{\hat{B}_{B_s}} 
= 266\pm 18 \textrm{ GeV}$. For other inputs, we use the values given in Appendix~\ref{inputs}. 

\begin{figure}
\includegraphics[scale=0.75]{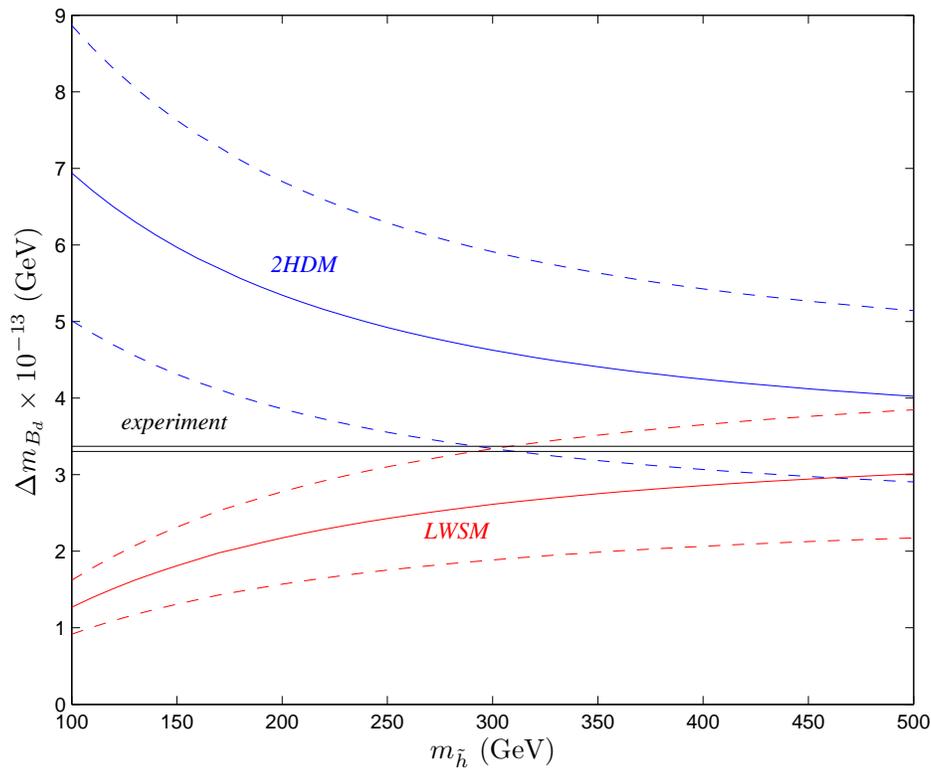}
\caption{$B_d-\bar{B}_d$ mass splitting predictions in the 2HDM of type-II with $\tan\beta=1$ and 
in the LWSM, as functions of the charged Higgs mass.  The solid curved lines give the central values
of the theoretical predictions, while the dashed lines delimit their $2\sigma$ error bands.  The solid
horizontal lines give the $2\sigma$ experimentally allowed region.}
\label{fig:Bd}
\end{figure}

\begin{figure}
\includegraphics[width=120mm,angle=0]{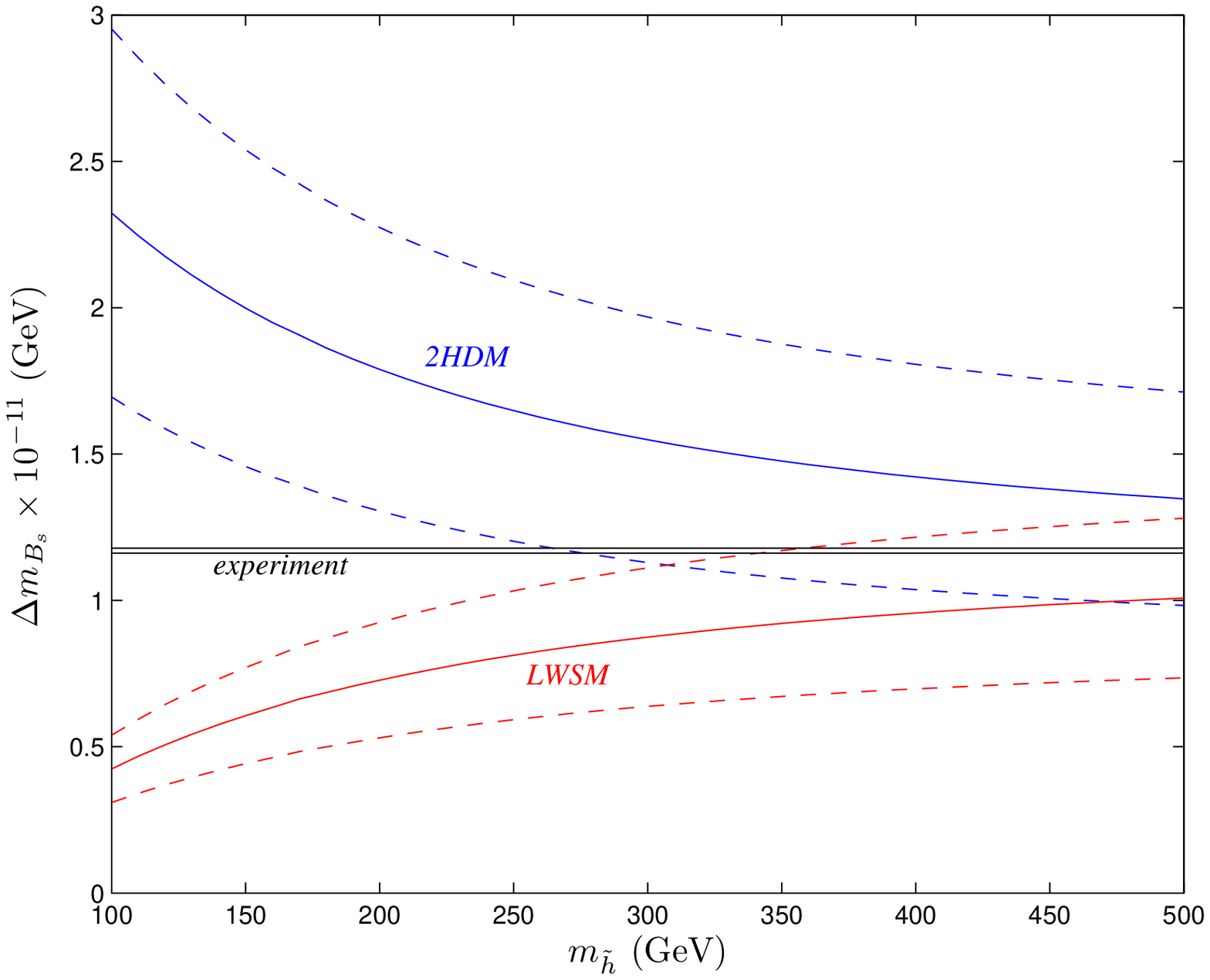}
\caption{$B_s-\bar{B}_s$ mass splitting predictions in the 2HDM of type-II with $\tan\beta=1$ and 
in the LWSM, as functions of the charged Higgs mass.  The solid curved lines give the central values
of the theoretical predictions, while the dashed lines delimit their $2\sigma$ error bands.  The solid
horizontal lines give the $2\sigma$ experimentally allowed region.}
\label{fig:Bs}
\end{figure}

There is an immediate question on the proper choice for the CKM matrix elements required to produce 
a theoretical prediction.  These elements are extracted, in part, from global fits that include the 
very process that is affected by the new physics.  The simplest approach (and one that seems standard in
the literature) is to use the best global fit values for the CKM elements in the SM.  One then requires 
that the theoretical prediction for the process of interest not deviate by more than a prescribed amount 
(approximately two standard deviations) from the experimental value.  This approach is sensible 
because the global SM fit of CKM elements is consistent with the experimental data.  More precisely,
our bounds are determined using a $\chi^2$ test:
\be
\chi^2_i=\f{(\mathcal O_{i,LWSM}-\mathcal O_{i,expt})^2}{\sigma^2_{i}},
\ee 
where $\mathcal O_{i,LWSM}$ is LWSM prediction for a particular process, $\mathcal O_{i,expt}$ 
is the related experimental result and $\sigma_{i}$ incorporates the error coming from both the theoretical 
prediction and the experimental result. We require that $\chi^2_i$ can not exceed $3.84$ in order to 
determine the 95\% C.L. bound on the charged Higgs mass.  Theoretical uncertainties (described above)
are added in quadrature with the experimental error in applying the $\chi^2$ test.  We assume the
experimental values~\cite{pdg08} $\Delta m_{B_d}= ( 3.337 \pm 0.033 ) \times 10^{-10} \textrm{ MeV}$ 
and $\Delta m_{B_s}=( 117.0 \pm 0.8 ) \times 10^{-10} \textrm{ MeV}$.   The results of this analysis
are shown in Figs.~\ref{fig:Bd} and \ref{fig:Bs}.

With the main theoretical error originating from the bag factors, we find that
the total error in both the $B_d$ and $B_s$ systems is well approximated by
\be
\sigma = 0.14 \times \Delta m_{B_{LWSM}},
\ee
which reflects that the experimental uncertainty is negligible compared to the theoretical one.
Applying our $\chi^2$ test to $B_d\bar{B}_d$ mixing, we find that the mass of LW charged 
Higgs boson is bounded by
\be \label{eq:bound1}
m_{\tilde{h}} > 303 \textrm{ GeV} \;\;\;\;\;\;\; (95\% \textrm{ C.L.}) \,,
\ee
while from $B_s\bar{B}_s$ mixing, 
\be \label{eq:bound2}
m_{\tilde{h}} > 354 \textrm{ GeV} \;\;\;\;\;\;\; (95\% \textrm{ C.L.}).
\ee
Note that the bound from $B_d\bar{B}_d$ is almost identical in the type-II 2HDM with $\tan\beta=1$, 
where one would find $m_{h^\pm} > 308$~GeV using the same method of analysis.  However, this is 
purely coincidental.  If the theoretical uncertainties were reduced by a factor of $2$ one would 
find that these particular bounds change to $446$~GeV (LWSM) and $618$~GeV (2HDM, $\tan\beta=1$), 
consistent with our earlier comment that the predictions in these two theories are qualitatively 
different.  

\subsection{$B \rightarrow X_s \gamma$}

\begin{figure}
\centering
\includegraphics[width=75mm,angle=0]{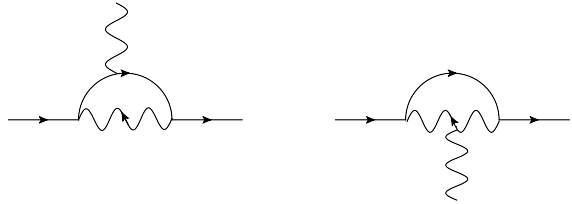}
\caption{SM contributions to $B \rightarrow X_s \gamma$. Wavy lines with (without) arrows represent 
$W$ bosons (photons), and solid lines represent quarks.}
\label{fig:bs1}
\end{figure}

\begin{figure}
\centering
\includegraphics[width=75mm,angle=0]{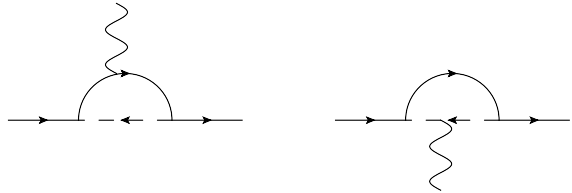}
\caption{Charged Higgs contributions to $B \rightarrow X_s \gamma$. Dashed lines represent Higgs fields, 
solid lines represent quarks and wavy lines represent photons.}
\label{fig:bs2}
\end{figure}

The bounds presented in Eqs.~(\ref{eq:bound1}) and (\ref{eq:bound2}) were limited by the
theoretical uncertainties in the lattice calculations of the bag factors.  We now consider
an observable that does not have this uncertainty, namely the ratio of the inclusive 
$B$ decay width $\Gamma(B \rightarrow X_s \gamma)$ to $\Gamma(B \rightarrow X_c e \overline{\nu}_e)$.
Standard model diagrams for $B \rightarrow X_s \gamma$ are shown in Fig.~\ref{fig:bs1} and new physics 
diagrams in Fig.~\ref{fig:bs2}.  In the LWSM, the diagrams of Fig.~\ref{fig:bs2} differ by an
overall sign relative to those of a type-II 2HDM with $\tan\beta=1$.

\begin{figure}
\includegraphics[scale=0.75]{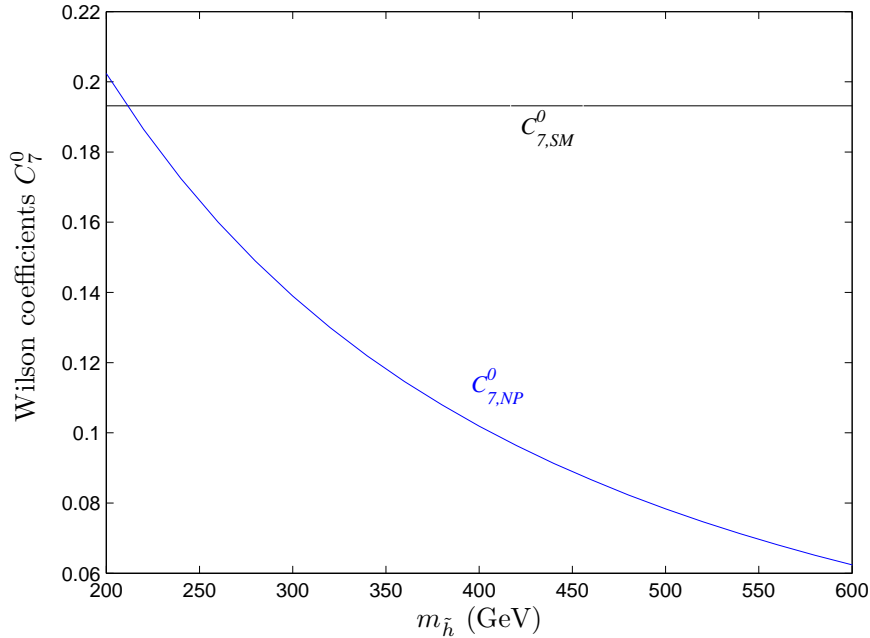}
\caption{Contributions to the Wilson coefficient $C^0_7$ from the Standard Model (SM) and from
New Physics (NP). The curve labelled $C^0_{7,NP}$ corresponds to the 2HDM with $\tan\beta=1$ or
$-C^0_{7,NP}$ for the LWSM.}
\label{fig:Wilson}
\end{figure}

As in our discussion of $B \bar B$ mixing, we first consider the leading-order contributions
to $B \rightarrow X_s \gamma$, evaluated at the matching scale $m_W$, to gain some insight on the 
effect of new physics. The branching fraction is given by~\cite{ogw}
\be
\mathcal{B}(B\rightarrow X_s\gamma) = 
\mathcal B(B \rightarrow X_ce\bar\nu_e) \left|\f{V_{ts}^*V_{tb}}{V_{cb}}\right|^2 
\f{6\alpha_{em}}{\pi f(m^2_c/m^2_b)} \left|C^0_{7,SM}+C^0_{7,NP}\right|^2,
\ee
where $C^0_7$ are Wilson coefficients.  In the 2HDM of type II, these are given by~\cite{ogw}
\be
C^0_{7,SM}=\f{x}{24}\lf[\f{-8x^3+3x^2+12x-7+(18x^2-12x)\ln(x)}{(x-1)^4}\rh] \, , 
\ee
\be
C^0_{7,NP}=\f{1}{3}\cot^2(\beta)\; C^0_{7,SM}(x\rightarrow y) 
+ \f{1}{12}y\lf[\f{-5y^2+y-3+(6y-4)\ln(y)}{(y-1)^3}\rh] \, ,
\ee
where $x=\f{m_t^2}{m_W^2}$ and $y=\f{m_t^2}{m_H^2}$, while in the LWSM
\be
C^0_{7,NP}=-\f{1}{3} C^0_{7,SM}(x\rightarrow y) - \f{1}{12}y\lf[\f{-5y^2+y-3+(6y-4)\ln(y)}{(y-1)^3}\rh] \, .
\ee
The function $f$ is a phase space suppression factor from the semileptonic decay rate 
\be
f(z)=1-8z_0+8z_0^3-z_0^4-12 z_0 ^2\ln z_0 \, ,
\ee
where $z_0=m_c^2/m_b^2$.  A plot of Wilson coefficients $C^0_7$ is provided in 
Fig.~\ref{fig:Wilson}.  This figure indicates that the new physics gives a positive contribution to 
branching fraction in the 2HDM, but a negative one in the LWSM, leading to a difference in the bound 
on the mass of the charged Higgs.

\begin{figure}
\includegraphics[scale=0.75]{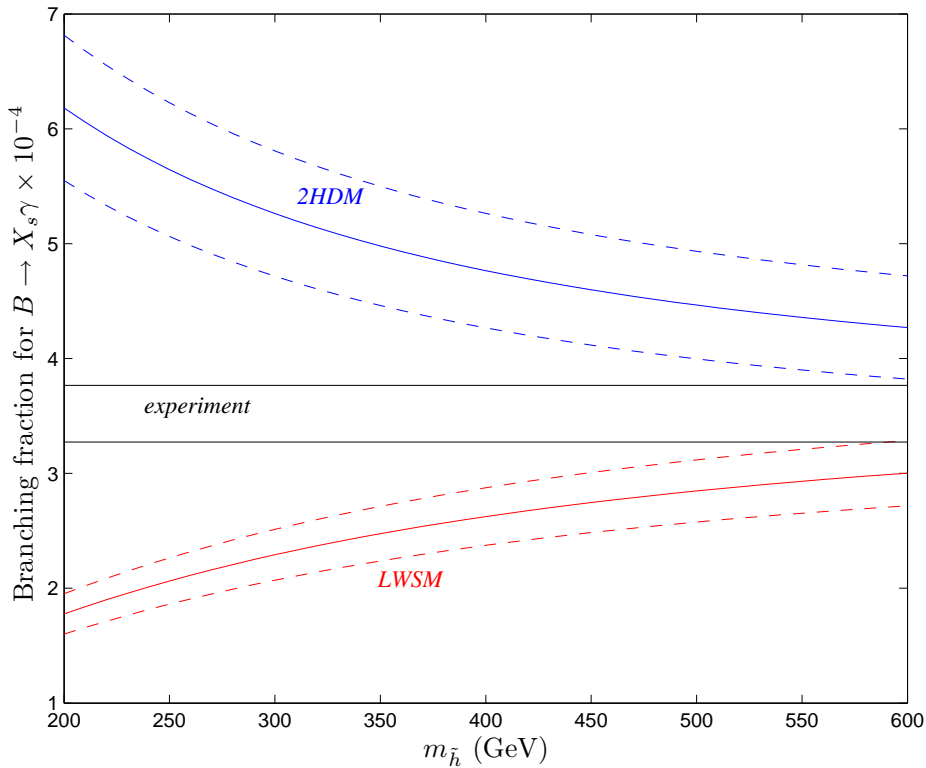}
\caption{Predictions for the branching fraction $\mathcal{B} (B\rightarrow X_s \gamma)$ 
in the type-II 2HDM with $\tan\beta=1$ and in the LWSM as functions of the charged Higgs mass.
The solid curved lines give the central values of the theoretical predictions, while the dashed 
lines delimit their $1\sigma$ error bands.  The solid horizontal lines give the $1\sigma$ 
experimentally allowed region.}
\label{fig:Bdecay}
\end{figure}

Following the approach used in the previous subsection, we obtain more accurate numerical 
bounds by modifying the 2HDM results that include NLO QCD corrections.  These expressions 
cannot be summarized in a few lines, and are taken from Ref.~\cite{ciuch}. All relevant input 
parameters are given in Appendix~\ref{inputs}.  As discussed in Ref.~\cite{ciuch}, the 
theoretical uncertainty comes from the error bars on physical input parameters as well as the 
choice of a number of renormalization scales.  The scales $\mu_b$ and $\bar \mu_b$
defined in Ref.~\cite{ciuch} refer to the $B$ meson renormalization scale in the
$b \rightarrow X_s \gamma$ and $b \rightarrow X_c e \overline{\nu}_e$ amplitudes, respectively, while 
$\mu_W$ is the scale at which the full theory is matched to the low-energy effective theory.
The theoretical error is determined, in part, by varying these scales from half to twice of 
their central values.  Errors coming from varying these scales and those originating from input 
parameters uncertainties are added in quadrature to obtain a total theoretical error.  The experimentally
allowed range is given by $\mathcal{B} (B\rightarrow X_s \gamma) 
= (3.52\pm 0.23 \pm0.09)\times 10^{-4}$~\cite{Barberio:2008fa}. Our results for the LWSM and 
for the type-II 2HDM with $\tan\beta=1$ are plotted in Fig.~\ref{fig:Bdecay}.  Note that the 
theoretical predictions asymptote at large values of the charged Higgs mass to the Standard Model
prediction $\mathcal{B} (B\rightarrow X_s \gamma) = (3.60\pm 0.36 )\times 10^{-4}$.

The bound on the charged Higgs mass is obtained by using the $\chi^2$ test described in 
the previous subsection.  Including the NLO QCD corrections, the bound in the LWSM is
\be \label{eq:constBsgamma}
m_{\tilde{h}} > 463 \textrm{ GeV} \;\;\;\;\;\;\; (95\% \textrm{ C.L.}) .
\ee
In Fig.~\ref{fig:Bdecay} we display $1\sigma$ theoretical and experimental error
bands.  Note that for approximately equal errors $\sigma_0$,  the separation between 
the experimental and theoretical central values that corresponds to a $\chi^2$ of $3.84$ 
is $\sim 2.8 \,\sigma_0$.  Using this observation, one can confirm that the bound in 
Eq.~(\ref{eq:constBsgamma}) and Fig.~\ref{fig:Bdecay} are consistent.

\subsection{$R_b$ from $Z$ Decay}

\begin{figure}
\centering
\includegraphics[width=75mm,angle=0]{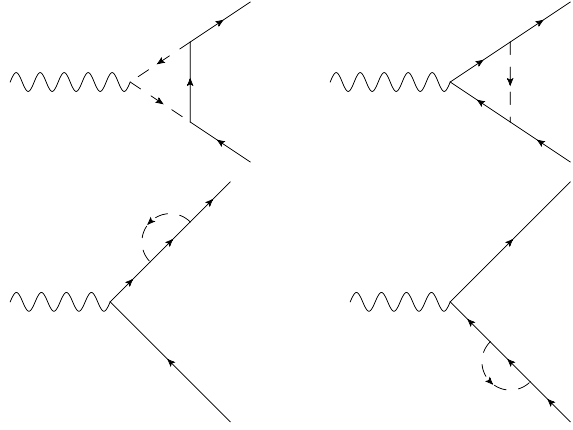}
\caption{Charged Higgs contributions to $Z \rightarrow b \bar b$ decay. Wavy lines represent 
Z bosons, solid lines represent quarks and dashed lines represent charged Higgs fields.}
\label{fig:Rb}
\end{figure}

\begin{figure}
\centering
\includegraphics[width=75mm,angle=0]{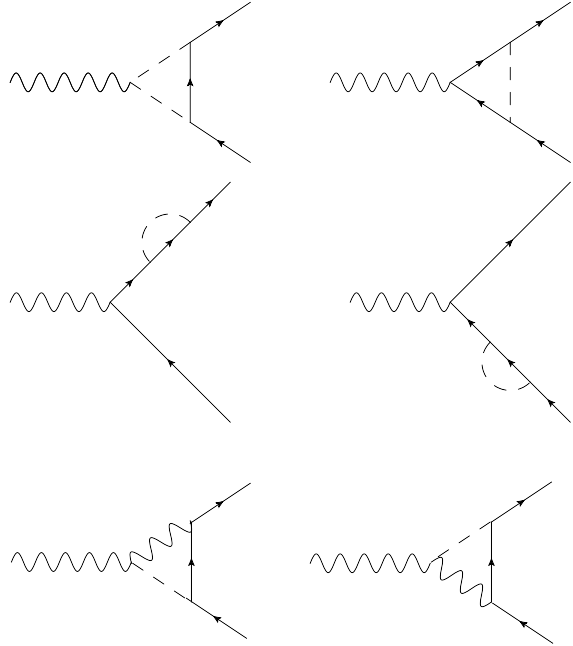}
\caption{Neutral Higgs contributions to $Z \rightarrow b \bar b$ decay. Wavy lines represent 
Z bosons, solid lines represent quarks and dashed lines represent neutral Higgs fields (scalars and
pseudoscalar).}
\label{fig:Rbneutral}
\end{figure}

The bounds that we have obtained thus far have followed from the consideration of flavor
changing neutral current processes.  It is also interesting to consider the effect of
new Higgs physics on the flavor-conserving $Zb\overline{b}$ coupling, which is measured
to high precision.  Here we focus on the observable $R_b \equiv \Gamma(Z \rightarrow b\bar{b})/
\Gamma(Z \rightarrow \mbox{ hadrons})$.  The charged-Higgs diagrams that contribute to this
process are shown in Fig.~\ref{fig:Rb}, while the neutral Higgs diagrams are shown in
Fig.~\ref{fig:Rbneutral}.  We will argue that the top-Yukawa-enhanced charged Higgs diagrams
are the only one necessary to obtain a numerically accurate result, and that the LWSM prediction
can be obtained, as before, by modifying the 2HD model result, which can be found, in this case, in
Ref.~\cite{howie}.

Let us first consider the possible contribution from the neutral Higgs fields. The interaction 
Lagrangian involving the neutral Higgs fields and quarks is given by
\be \label{eq:Hff}
\mathcal L_{Hff} = - \sum_{f} \f{m_f}{v} \lf\{(\cosh\theta-\sinh\theta)h_0
+(\sinh\theta-\cosh\theta)\tilde h_0+i\tilde P\rh\}\bar f_Rf_L + \textrm{ h.c.},
\ee
where sum extends over all fermions in the SM. The interactions between the $Z$-boson and the 
neutral Higgs fields are given by:
\be
\mathcal L_{ZHH}=\f{g_2}{2\cos\theta_W}\lf[\sinh\theta\lf( 
h_0\partial\mu \tilde P-\tilde P\partial_\mu h_0\rh) + \cosh\theta\lf(
\tilde h_0\partial\mu \tilde P - \tilde P\partial_\mu\tilde h_0 \rh)\rh]Z^\mu,
\ee
and
\be \label{eq:neutralHZZ} 
\mathcal L_{ZZH} = \f{g_2 m_Z}{2\cos\theta_W}Z_\mu^2 \left(\cosh\theta h_0+\sinh\theta \tilde h_0\right).
\ee
Clearly the neutral Higgs amplitudes that contribute to $R_b$ are smaller than those of the 
charged Higgs by $m_b^2/m_t^2 \sim 6 \times 10^{-4}$ and can be neglected provided that there
is no compensating enhancement from any other source.  One might worry in the LWSM that the
factors of $\cosh\theta$ and $\sinh\theta$, which can be arbitrarily large, might provide such
an effect.  This fear, however, is unfounded.  Every relevant $Z \rightarrow b \overline{b}$ amplitude
could have been computed in the HD formulation of the LWSM where there are clearly no couplings that
are becoming large.  The particle mass eigenvalues and the hyperbolic functions of mixing angles in
the LW form of the theory must therefore combine in physical amplitudes so that the same outcome
is obtained.  As a pedagogical example, one can consider the diagrams with fermion wave function 
renormalization due to a Higgs field loop shown in Fig.~\ref{fig:Rbneutral}. The product of the scalar 
propagators and fermion couplings is proportional to
\begin{equation} \label{eq:recoverhd}
(\cosh\theta-\sinh\theta)^2 \left[ \frac{i}{p^2-m_{h_0}^2} - \frac{i}{p^2-m_{\tilde{h}_0}^2}\right]  \, ,
\end{equation}
where $p$ is the internal momentum on the scalar line.  From Eq.~(\ref{eq:eigenmass}) it follows that
\begin{eqnarray}
&& m^2_{h_0}+m^2_{\tilde{h}_0} = m^2_{\tilde{h}} \,, \nonumber \\
&& m^2_{h_0}\, m^2_{\tilde{h}_0} = m^2_{h}\, m^2_{\tilde{h}} \, \nonumber \\
&& m^2_{h_0}-m^2_{\tilde{h}_0} = - m^2_{h} \sqrt{1-4 m^2_{h}/m^2_{\tilde{h}}} \,,
\end{eqnarray}
and using Eq.~(\ref{tanhform})
\begin{equation}
(\cosh\theta-\sinh\theta)^2 = \frac{1}{\sqrt{1-4 m^2_{h}/m^2_{\tilde{h}}}} \,
\end{equation}
from which one can easily show that Eq.~(\ref{eq:recoverhd}) can be rewritten
\begin{equation}
\frac{i}{p^2 - m_h^2 - p^4/m_{\tilde{h}}^2} \,,
\end{equation}
which has no singular behavior as either hyperbolic function becomes infinite.  It is therefore
safe to drop $m_b^2/m_t^2$ suppressed effects, as is usually done in conventional 2HD models.

\begin{figure}
\includegraphics[width=120mm,angle=0]{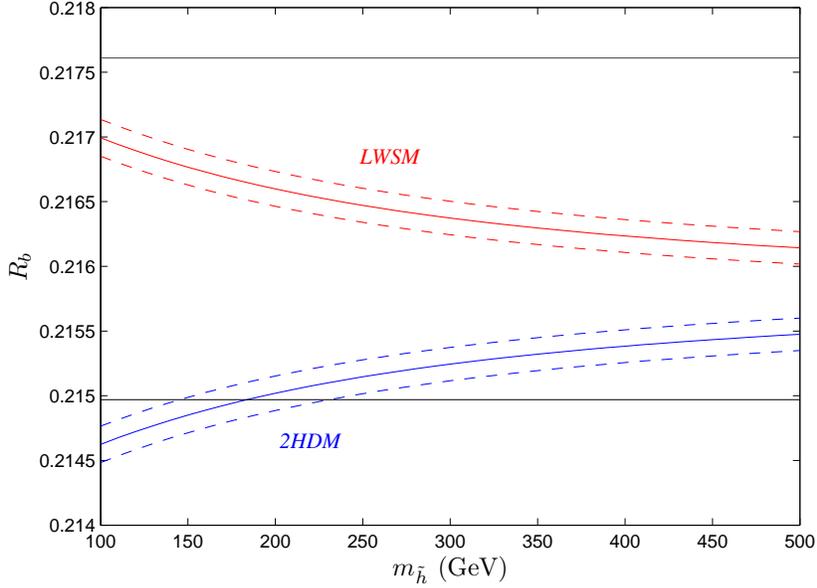}
\caption{$R_b$ predictions for the 2HDM of type-II with $\tan\beta=1$ and for the LWSM as functions 
of the charged Higgs mass. The solid curved lines give the central values
of the theoretical predictions, while the dashed lines delimit their $2\sigma$ error bands.  The solid
horizontal lines give the $2\sigma$ experimentally allowed region.}
\label{fig:Rbplot}
\end{figure}

In the type-II 2HDM, the corrections to the left- and right-handed $b$-quark couplings to the $Z$ boson are
given by~\cite{howie}
\beq \label{eq:RbHD}
\delta g^L&=&\f{1}{32\pi^2}\lf(\f{g_2 m_t}{\sqrt{2}m_W}\cot\beta\rh)^2\f{e}{s_W c_W}\lf[\f{R}{R-1}
-\f{R\ln R}{(R-1)^2}\rh] \, , \nonumber\\
\delta g^R&=&-\f{1}{32\pi^2}\lf(\f{g_2 m_b}{\sqrt{2}m_W}\tan\beta\rh)^2\f{e}{s_W c_W}\lf[\f{R}{R-1}
-\f{R\ln R}{(R-1)^2}\rh] \, ,
\eeq
where $R=m_t^2/m_{\tilde h}^2$. For $\tan\beta\simeq 1$, the correction $\delta g^R$ is negligible since 
it is $O(m_b^2/m_t^2)$ smaller than $\delta g^L$.   Therefore, the leading correction to the $Zb\overline{b}$
vertex in the LWSM is given by
\be \label{eq:RbLW}
\delta g^L=-\f{1}{32\pi^2}\lf(\f{g_2 m_t}{\sqrt{2}m_W}\rh)^2\f{e}{s_W c_W}\lf[\f{R}{R-1}
-\f{R\ln R}{(R-1)^2}\rh].
\ee
In the Standard Model, the best global fit value for $R_b$ is $0.21629 \pm 0.00066$, while the Standard Model
prediction is $0.21584 \pm 0.00006$~\cite{pdg08};  the LWSM gives a positive contribution to 
$R_b$ which helps reconcile the central values.  The results in a type-II 2HDM with $\tan\beta=1$ 
and in the LWSM are plotted in Fig.~\ref{fig:Rbplot}.  Since the LWSM correction pushes $R_b$ toward its 
experimental value, we do not obtain any bound on the charged Higgs mass from this process.

\section{Constraining the Neutral Sector}\label{sec:four}

It is worth recalling that the parameter we have been bounding, $m_{\tilde{h}}$, determines both
the charged and pseudoscalar LW Higgs masses.  Bounds as high as those obtained in the previous section
easily supercede those of direct collider searches for charged Higgs bosons in type-II
2HD models, which are typically below $80$~GeV~\cite{pdg08}.  These bounds should apply to
the LWSM since the overall sign-flips in the tree-level diagrams that determine the 
direct production of charged LW Higgs do not affect the production rate.  

We now consider what can be said about the allowed parameter space for the remaining, scalar LW Higgs 
fields.  We determine the allowed region in the $m_{h_0}$-$m_{\tilde{h}_0}$ mass eigenvalue plane, the most
convenient parameter space for comparison to future collider searches.  From
Eq.~(\ref{eq:massrelation}), which relates the masses of the charged and neutral scalars, and
using our strongest bound on $m_{\tilde{h}}$ from $B \rightarrow X_s \gamma $, we have
\be \label{eq:chargedconstraint}
m_{h_0}^2+m_{\tilde h_0}^2 > (463~\textrm{ GeV})^2 \, .
\ee 
As noted earlier, Eq.~(\ref{eq:eigenmass}) implies that 
\be \label{byconst}
m_{\tilde{h}_0} > m_{h_0} \,.
\ee
Together, Eqs.~(\ref{eq:chargedconstraint}) and (\ref{byconst}) lead to a lower bound
on the $\tilde{h}_0$ mass
\be\label{eq:mhtb}
m_{\tilde h_0}> 327 \textrm{ GeV}.
\ee
Finally, searches for the Higgs at LEP allows us to determine a direct bound on $m_{h_0}$. At LEP,
the Higgs is produced via the Higgstrahlung process $e^+e^- \rightarrow Z^* \rightarrow h_0 Z$, which 
involves the vertex given in Eq.~(\ref{eq:neutralHZZ}).  Taking into account the bound on $m_{\tilde h_0}$
in Eq.~(\ref{eq:mhtb}), and the kinematical range accessible at LEP we conclude that the $h_0$ and
$\tilde{h}_0$ are well separated and we can neglect effects due to the $\tilde{h}_0$ which may
contribute to the same final states ({\em e.g.}, $e^+e^- \rightarrow Z b \overline{b}$).  The $h_0 Z Z$
coupling differs from the Standard Model by a factor of $\cosh\theta$, but this exceeds the
Standard Model value $\cosh\theta=1$ by no more than $1$\% in the LEP search.  In addition, a factor
of $(\cosh\theta-\sinh\theta)$ that appears in the Higgs couplings to fermions [{\em c.f.} Eq.~(\ref{eq:Hff})]
does so universally, and even enhances the branching fraction to these modes compared to the 
far subleading three-body decay channels.  Thus, referring to Ref.~\cite{Barate:2003sz}, we expect 
the LEP lower bound
\be \label{eq:neutralconstraint} 
m_{h_0}> 114 \textrm{ GeV}.
\ee 
to be approximately valid, and at the very least, to be slightly below the actual bound that
could have been obtained if LEP did a dedicated Lee-Wick analysis.

\begin{figure}
\includegraphics[scale=0.75]{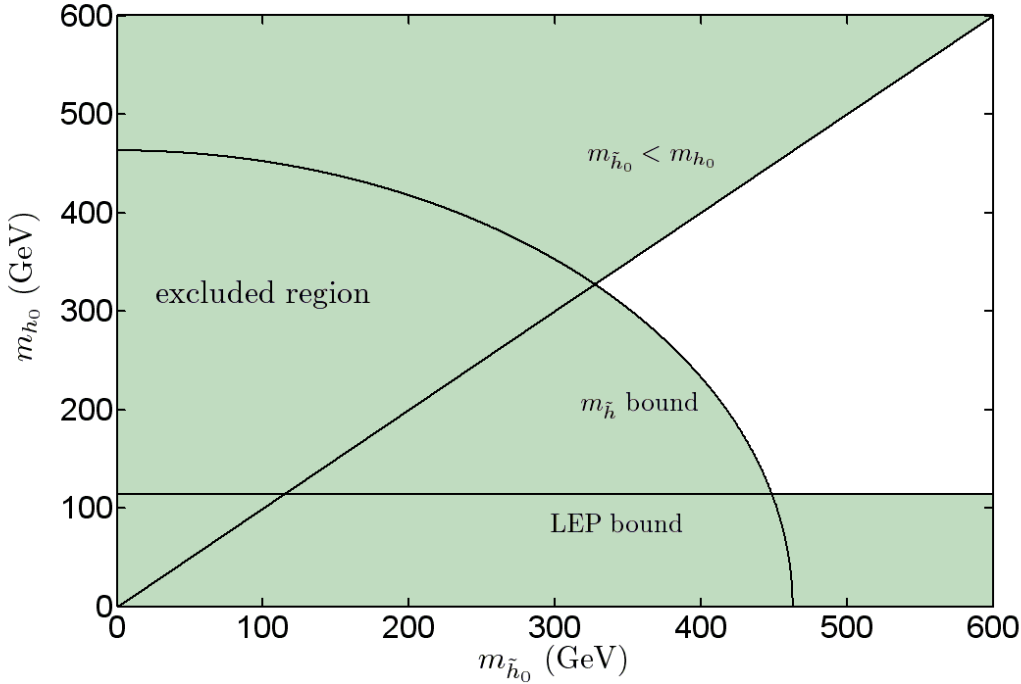}
\caption{Bounds on the $m_{h_0}$-$m_{\tilde{h}_0}$ plane. The shaded region is excluded}
\label{fig:Exclusion}
\end{figure}

We plot the excluded regions in the $m_{h_0}$-$m_{\tilde{h}_0}$ plane in Fig.~\ref{fig:Exclusion}.
The shaded quarter circle represents the indirect constraint obtained from our charged Higgs
bound Eq.~(\ref{eq:chargedconstraint}), the horizontal line represents the LEP bound 
Eq.~(\ref{eq:neutralconstraint}), and the diagonal line indicates where Eq.~(\ref{byconst}) holds.

\section{Conclusions}\label{sec:concl}

We have studied the possibility that the Higgs sector of the Lee-Wick Standard Model 
is lighter than the remaining Lee-Wick particle content, a possibility that is consistent
with Higgs sector naturalness and precision electroweak constraints.  The effective theory
is a two-Higgs doublet model in which one doublet has wrong-sign kinetic and mass terms. 
By considering $B_d$-$\overline{B}_d$ and $B_s$-$\overline{B}_s$ mixing and the decay 
$b \rightarrow X_s \gamma$, we obtained the bounds $m_{\tilde{h}}>303$~GeV, $354$~GeV and
$463$~GeV, respectively, where $m_{\tilde{h}}$ is the mass of the Lee-Wick charged Higgs $\tilde{h}^\pm$
and also the mass of the Lee-Wick pseudoscalar $\tilde{P}$.  We then studied the decay 
$Z \rightarrow b \overline{b}$ and found that the Lee-Wick Higgs corrections to the Stardard Model 
prediction provided better agreement with the experimental data and that no additional bound 
could be obtained.  Finally, we argued the LEP search for the Higgs boson implies the 
bound $m_{h_0}>114$~GeV on the ordinary Higgs scalar in the Lee-Wick Standard Model.  Study 
of the allowed regions of the $m_{h_0}$-$m_{\tilde{h}_0}$ plane led us to an absolute lower 
bound on the Lee-Wick neutral scalar mass $m_{\tilde{h}_0}>327$~GeV.  Interestingly, all our 
bounds indicate that the Lee-Wick Higgs sector could be within the kinematic reach of the LHC.

Clearly, a vast literature exists on two-Higgs doublet model constraints and collider 
signals --- other processes surely exist to which the present analysis could be extended.
The work reported here is intended as a first step in exploring a two-Higgs doublet model 
of an unconventional sort, and a reminder that Lee-Wick physics can conceivably lurk well below 
$1$~TeV.

\acknowledgments
We thank the NSF for support under Grant Nos.\ PHY-0456525 and PHY-0757481.

\appendix
\section{Numerical Inputs} \label{inputs} 

Unless referenced otherwise below, the numerical inputs used in our analysis are 
taken from Particle Data Group~\cite{pdg08}: 
\begin{center}
\begin{tabular}{lcl}
\hline\hline
\multicolumn{3}{c}{Quarks, gauge bosons and B mesons masses} \\
\hline
$m_t = 171.2\pm 2.1 \textrm{ GeV} $ & \qquad\qquad &$m_W=80.398\pm 0.025 \textrm{ GeV}$ \\
$\bar m_b (\bar m_b) = 4.2^{+0.17}_{-0.07} \textrm{ GeV}$& & $m_Z=91.1876 \pm 0.0021 \textrm{ GeV}$ \\
$\bar m_c (\bar m_c) = 1.27^{+0.07}_{-0.11} \textrm{ GeV}$& & $m_{B_d}=5279.53 \pm 0.33 \textrm{ MeV}$ \\
$m_s = 104^{+26}_{-34} \textrm{ MeV}$& & $m_{B_s}=5366.3 \pm 0.6 \textrm{ MeV}$ \\
\hline\hline
\end{tabular}
\end{center}
\begin{center}
\begin{tabular}{lcl}
\hline\hline
\multicolumn{3}{c}{Wolfenstein parameters} \\
\hline
$\lambda = 0.2257^{+0.0009}_{-0.0010}$ &\qquad\qquad & $A = 0.814^{+0.021}_{-0.022}$ \\
$\bar{\rho} = 0.135^{+0.031}_{-0.016}$ && $\bar{\eta} = 0.349^{+0.015}_{-0.017}$\\
\hline\hline
\end{tabular}
\end{center}
(The relationship between the Wolfenstein parameters and the $V_{ij}$ that preserves
unitarity to all orders in $\lambda$ can be found in Ref.~\cite{pdg08}.)
\begin{center}
\begin{tabular}{lcl} \hline\hline
\multicolumn{3}{c}{Other parameters} \\
\hline
$G_F=1.16637\times10^{-5} \textrm{ GeV}^{-2}$& \qquad\qquad &  $\alpha_{em}^{-1}=137.035999679$ \\
$\alpha_s(m_Z)=0.1176\pm0.0020$&&$s_W^2=0.23119 \pm 0.00014$\\
$f_B \sqrt{\hat{B}_{B_d}} = 216\pm 15 \textrm{ GeV}$~\cite{Gamiz:2009ku} &&
$\mathcal B(B \rightarrow X_ce\bar\nu_e)
=(10.74 \pm 0.16)\%$~\cite{Barberio:2008fa}\\
$f_B \sqrt{\hat{B}_{B_s}} = 266\pm 18 \textrm{ GeV}$~\cite{Gamiz:2009ku} && 
$(m_B^2-m^2_{B^*})/4=0.12 \textrm{ GeV}^2$\\
\hline\hline
\end{tabular}
\end{center}

Note that the top quark mass is the pole mass, while the remaining quark masses are running masses in
the $\overline{MS}$ scheme.

\end{document}